\documentclass{article}

\usepackage{xurl}
\usepackage{acronym}
\usepackage{amsmath}
\usepackage{amssymb}
\usepackage{authblk}
\usepackage{cleveref}
\usepackage{graphicx}
\usepackage{siunitx}
\usepackage{xspace}

\usepackage{fullpage}

\newacro{ABF}{artificial bacterial flagellum}
\newacroplural{ABF}{artificial bacterial flagella}
\newacro{DPD}{dissipative particle dynamics}
\newacro{GPU}{graphics processing unit}
\newacro{GCP}{Google Cloud Platform}
\newacro{HPC}{high performance computing}
\newacro{MD}{molecular dynamics}
\newacro{RBC}{red blood cell}
\newacro{RCB}{recursive coordinate bisectioning}
\newacro{RL}{reinforcement learning}

\newcommand{\Ht}{\mathrm{Ht}}
\newcommand{\Lmp}{LAMMPS\xspace}
\newcommand{\Mir}{Mirheo\xspace}
\newcommand{\Udx}{uDeviceX\xspace}
\newcommand{\key}[1]{\raisebox{0.5pt}{\protect\includegraphics{figures/keys/#1}}\,}

\title{Scalable, Cloud-Based Simulations of Blood Flow and Targeted Drug Delivery\\ in Retinal Capillaries}

\author[1]{Lucas Amoudruz\thanks{Equal contributions.}}
\author[1]{Sergey Litvinov\thanks{Equal contributions.}}
\author[2]{Riccardo Murri}
\author[3]{Volker Eyrich}
\author[2]{Jens Zudrop}
\author[2]{Costas Bekas}
\author[1]{Petros Koumoutsakos\thanks{Corresponding author.
    \textit{Email:} petros@seas.harvard.edu}}

\affil[1]{Computational Science and Engineering Laboratory, School of Engineering and Applied Sciences, Harvard University, Cambridge, MA 02138, United States.}
\affil[2]{Citadel Securities, 200 S. Biscayne Boulevard, Miami, FL 33131, United States.}
\affil[3]{Google, Mountain View, CA, United States.}

\date{}

\begin{document}

\maketitle

\begin{abstract}
  We investigate the capabilities of cloud computing for large-scale,tightly-coupled simulations of biological fluids in complex geometries, traditionally performed in supercomputing centers.
  We demonstrate scalable and efficient simulations in the public cloud.
  We perform meso-scale simulations of blood flow in image-reconstructed capillaries, and examine targeted drug delivery by artificial bacterial flagella (ABFs).
  The simulations deploy dissipative particle dynamics (DPD) with two software frameworks, \Mir (developed by our team) and LAMMPS. \Mir exhibits remarkable weak scalability for up to 512 GPUs.
  Similarly, LAMMPS demonstrated excellent weak scalability for pure solvent as well as for blood suspensions and ABFs in reconstructed retinal capillaries.
  In particular, LAMMPS maintained weak scaling above 90\% on the cloud for up to 2,000 cores.
  Our findings demonstrate that cloud computing can support tightly coupled, large-scale scientific simulations with competitive performance.
\end{abstract}

\section{Introduction}

Computing has become the third pillar of scientific inquiry and engineering design, along with theory and experimentation.
We distinguish two major computing modalities: large-scale simulations, where all computational nodes are dedicated to solving one particular problem, and high-throughput, which involves many smaller simulations run independently across different compute nodes.
In the former case, communication within and between nodes is essential, whereas in the latter case, communication between nodes is reduced.
High-throughput computing has become indispensable for optimization and uncertainty quantification in areas ranging from manufacturing to automotive design.

To date, the vast majority of computing, in particular for large-scale simulations, has relied on dedicated computational infrastructure such as local clusters, university computing centers, and national supercomputing facilities.
These types of infrastructure offer a unique combination of benefits in terms of accessibility, speed, cost, and service.
However, they often require long-term investments, which can impact short-term cost-efficiency, while access to them can be restricted or may not be readily available on demand.

In recent years, cloud computing has emerged as a disruptive trend by offering a flexible, scalable and potentially cost-efficient alternative to traditional \ac{HPC}.
Researchers initially evaluated cloud computing for relatively simple tasks, such as single-node computations~\cite{vecchiola2009}, or small-scale parallel runs~\cite{sadooghi2015}.
In early benchmarks, Iosup et al.~\cite{iosup2011performance} and Zhai et al.~\cite{zhai2011cloud} found that commercial cloud services (Amazon EC2, GoGrid, ElasticHosts, and Mosso) performed poorly compared to dedicated \ac{HPC} systems, largely due to limited interconnect speed and high latency.
These constraints were considered critical for tightly coupled simulations, where frequent communication between nodes is required.

However, this is not a major bottleneck for high-throughput workloads~\cite{martin2022korali,hadjidoukas2015pi4u,adams2020dakota}, and cloud computing has proven highly effective in this scenario~\cite{kutzner2022gromacs}.
Furthermore, recent effort has shown progress in enabling tightly coupled simulations on the cloud.
This includes microbenchmarks~\cite{fernandez2021evaluation}, tightly coupled simulations on four nodes~\cite{munhoz2023evaluating} and computational fluid dynamics at small to moderate scale~\cite{taylor2018enabling,mesnard2019reproducible,zaspel2011massively}.

Despite this progress, large-scale simulations of complex biological systems, such as blood flow at the microscale, remain dependent on traditional \ac{HPC} infrastructure.
These computing platforms offer high-bandwidth and low-latency interconnects but come with limitations: restricted and competitive access, limited scalability beyond allocated quotas, high operational overheads, and a lack of on-demand elasticity.
In contrast, modern cloud infrastructure offers elastic compute resources, user-based pricing, and global availability, potentially transforming current approaches in scientific computing.

In fact, other fields have already embraced cloud computing for tightly-coupled tasks.
For instance, weather forecasting models have been successfully migrated to the cloud, offering faster deployment cycles and reduced infrastructure costs~\cite{siuta2016viability,powers2021cloud}.
Similarly, industries such as automotive and aerospace have adopted cloud platforms to run commercial computational fluid dynamics codes~\cite{kumar2023performance,dancheva2023cloud,boiger2021massive}, citing improved flexibility and cost-efficiency as key drivers.

A previous study has demonstrated that the lattice–Boltzmann solver Palabos can scale efficiently for blood-flow simulations on public clouds~\cite{kotsalos2021palabos,Tang2023AWS}.
Yet, to our knowledge, no study has evaluated whether large-scale tightly coupled microscale blood simulations, using particle-based methods like \ac{DPD}, can be efficiently executed on public cloud infrastructure.

This work addresses this gap by investigating whether recent advances in cloud computing, such as enhanced networking, faster storage, and cost-optimized compute tiers, are sufficient to support demanding simulations of blood at sub-micron resolution.
These simulations are critical for biomedical applications, including targeted drug delivery, precision microsurgery, and microsensing.
The complexity of these simulations arises from the need to accurately predict the deformations of \acp{RBC} and their interaction with blood plasma.

We consider two software packages for this purpose.
Both programs represent blood as a set of \ac{RBC} membranes suspended in plasma, an incompressible viscous Newtonian fluid.
The membranes are discretized into a triangle mesh, while the plasma is represented by the \ac{DPD} method~\cite{hoogerbrugge1992simulating}.
The first program that we consider is \Lmp~\cite{thompson2022lammps}, a widely used \ac{MD} package parallelized across nodes with MPI.
Here, \Lmp was extended to support the addition of \acp{RBC} membrane model described in ref.~\cite{fedosov2010multiscale}, as well as \acp{ABF}~\cite{amoudruz2022phd,amoudruz2025optimalnavigationmagneticartificial}.

\Lmp is a widely used software package for particle simulations, known for its user-friendly design and relatively good performance~\cite{thompson2022lammps}.
It features an advanced spatial decomposition algorithm tailored for simulations on distributed memory hardware.
Its proven scalability on leading supercomputers further establishes its reputation for reliability and high performance in particle simulations.
\Lmp has been designed to optimize its critical performance kernels for diverse hardware environments, including \acp{GPU} and multi-threaded CPUs.
\Lmp achieves dynamic load balancing using the \ac{RCB} algorithm~\cite{bokhari1987partitioning}.
This algorithm plays a critical role in optimally distributing computational workloads, thus enhancing overall system efficiency by maintaining a balanced resource allocation.
The second program considered in this study is \Mir~\cite{alexeev2020mirheo}, a high-performance package for microfluidics simulations on multi-\ac{GPU} architectures.
\Mir includes a detailed and accurate model of blood that was extensively validated against experimental data in~\cite{amoudruz2023a}.
\Mir is an extension of \Udx~\cite{rossinelli2015silico}, a finalist for the Gordon Bell Prize in 2015, and was extensively optimized for the Nvidia P100 \acp{GPU} on the Cray XC50 Piz-Daint supercomputer at CSCS.
It demonstrated unprecedented time-to-solution and excellent weak scalability up to thousands of \acp{GPU} for blood simulations, as well as pure \ac{DPD} simulations on supercomputers.

\paragraph*{Contributions}
\begin{enumerate}
    \item We demonstrate that cloud computing can democratize access to tightly coupled large-scale scientific simulations.
    \item We show that cloud-based simulations can achieve competitive performance compared to traditional supercomputers, both with CPU and GPU programs.
    \item We introduce improvements to data structures and communication patterns in the widely used software \Lmp, significantly reducing all-to-all communication and message sizes between ranks.
    These enhancements enable excellent weak and strong scalability across thousands of cores.
    \item We show that cloud computing enables complex and tightly coupled simulations at scale.
    In this study, we focus on the example of sub-micron resolution blood flow simulations, but we emphasize that this methodology extends to many scientific applications.
\end{enumerate}

\section{Methods}
\label{sec:model}

Blood is modeled as a suspension of \acp{RBC} in plasma.
Each \ac{RBC} consists of a viscoelastic membrane that encloses the cytosol, a fluid six times more viscous than plasma.
The \ac{ABF} is treated as a rigid body with a magnetic moment, enabling its orientation and movement to be controlled by external magnetic fields.
Fluids inside and outside the \ac{RBC} are modeled using the \ac{DPD} method.
The \ac{RBC} membranes are discretized as particles that form a triangle mesh, while the \ac{ABF} consists of frozen \ac{DPD} particles moving as a rigid object.

\subsection{Dissipative Particle Dynamics}

Fluids are discretized into \ac{DPD} particles that have positions $\mathbf{r}_i$, velocities $\mathbf{v}_i$, and mass $m$, $i=1,2,\dots,N$.
The particles evolve according to Newton's laws of motion and interact with neighboring particles within a cut-off radius $r_c$.
These interactions are governed by pairwise forces, which consist of three distinct terms~\cite{hoogerbrugge1992simulating,espanol1995statistical},
\begin{equation*}
\begin{split}
  \mathbf{F}_{ij} = a w(r_{ij}) \mathbf{e}_{ij}
  - \gamma \left(\mathbf{e}_{ij} \cdot \mathbf{v}_{ij}\right)  w_D(r_{ij}) \mathbf{e}_{ij} \\
  + \sigma \xi_{ij} w_R(r_{ij}) \mathbf{e}_{ij},
\end{split}
\end{equation*}
where $\mathbf{v}_{ij} = \mathbf{v}_i - \mathbf{v}_j$, $\mathbf{r}_{ij} = \mathbf{r}_i - \mathbf{r}_j$, $r_{ij}$ = $\|\mathbf{r}_{ij}\|$ and
$\mathbf{e}_{ij} = \mathbf{r}_{ij} / r_{ij}$.
The coefficients $a$, $\gamma$, and $\sigma$ are the conservative, dissipative, and random force magnitudes, respectively.
Furthermore, we use the standard \ac{DPD} conservative kernel
\begin{equation*}
  w(r) =
  \begin{cases}
    1 - r/r_c, & r < r_c,\\
    0, & \text{otherwise}.
  \end{cases}
\end{equation*}
We set $w_D = w^{1/4}$ to obtain a high solvent viscosity~\cite{fan2006simulating}.
$w_R$ satisfies the fluctuation-dissipation theorem, $\sigma^2 = 2 \gamma k_BT$ and $w_D = w_R^2$~\cite{espanol1995statistical}, where $k_BT$ is the temperature of the system in energy units.
The \ac{DPD} method allows for the incorporation of complex wall boundaries.
Walls are modeled as a layer of frozen \ac{DPD} particles that interact with the other particles of the simulations via \ac{DPD} forces.
Furthermore, solvent particles are bounced-back from the surface of the walls.

\subsection{Red blood cell membranes}

We discretize the \ac{RBC} membrane into a triangulated mesh, composed of $N_v$ vertices and $N_s$ edges~\cite{Fedosov2010f}.
The elasticity of the spectrin cytoskeleton is decomposed into two terms: a shear energy modeled by in-plane elastic forces between vertices that share an edge and a stretch energy modeled by a local area potential.
The presence of the lipid bilayer is modeled through three different terms: (i) resistance to bending, incorporated through an energy potential whose magnitude depends on the angle between neighboring triangles, (ii) viscosity, modeled through viscous dissipation on the springs, and (iii) membrane incompressibility, represented by a global area penalization.
Finally, the incompressibility of the enclosed hemoglobin is represented by a volume penalization, as the solvent is not strictly modeled as an incompressible fluid.
The total potential energy of the \ac{RBC} membrane is therefore composed of four terms:
\begin{equation*}
U = U_s + U_b + U_A + U_V.
\end{equation*}
$U_s$ accounts for the energy of the elastic spectrin network of the \ac{RBC} membrane, modeled by an attractive worm-like chain potential and a repulsive potential such that a non-zero equilibrium spring length can be obtained,
\begin{equation*}
U_s = \sum_{j=1}^{N_{s}}
\left[
\frac {k_s l_{m} \left( 3x_j^2 - 2x_j^3 \right)}{4(1-x_j)} + \frac{k_{p_{j}}}{l_j}
\right],
\end{equation*}
where $k_{s}$ is a spring constant, $l_j$ is the length of spring $j$, $x_j=l_j/l_m$, $l_m$ is the maximum spring extension,
and $k_{p_j}$ is computed such that the total spring force on each spring is zero at equilibrium ($l_j = l_{0_j}$).
The bending energy term, $U_b$, represents the bending resistance of the lipid bilayer and is modeled as
\begin{equation*}
U_b = k_b \sum_{j=1}^{N_{s}} \left[ 1-\cos \theta_j \right],
\end{equation*}
where $k_b$ is a bending coefficient and $\theta_j$ is the angle between two adjacent triangles.
$U_A$ and $U_V$ represent the area and volume conservation constraints, respectively,
\begin{gather} \label{eq:penalization:A}
U_A = \frac{k_a (A-A_0)^2}{2A_0} + \sum_{j=1}^{N_{t}} \frac{k_d (A_j-A_{0_j})^2}{2A_{0_j}}, \\
U_V = \frac{k_v(V-V_0)^2}{2 V_0}, \label{eq:penalization:V}
\end{gather}
where $A_j$ and $A_{0_j}$ are the current and initial area of triangle $j$, $A$ and $A_0$ are the current and initial total membrane area, $V$ and $V_0$ are the current and initial volume enclosed by the membrane, and $k_a$, $k_d$, and $k_v$ are coefficients for the global area, local area, and volume, respectively.

The viscous dissipation of the membrane is modeled by adding a dissipative force term to the springs.
We use the membrane viscosity formulation presented in Fedosov et al.~\cite{Fedosov2010f} and set the non-central part of the force to zero ($\gamma^T=0$), as this term does not conserve angular momentum.
In this case, the dissipative force on a spring that connects vertices $i$ and $j$ is
\begin{equation*}
\mathbf{F}_{m, ij}^D = - \gamma^C \left( \mathbf{v}_{ij} \cdot \mathbf{e}_{ij}  \right) \mathbf{e}_{ij}
\end{equation*}
where $\mathbf{e}_{ij}$ is the unit vector along the membrane vertex centers, $\mathbf{e}_{ij} = \mathbf{r}_{ij} / \lVert\mathbf{r}_{ij}\rVert$, and $\mathbf{r}_{ij} = \mathbf{r}_i - \mathbf{r}_j$, with $\mathbf{r}_i$ being the positional vector of vertex $i$.

The parameters of the model were obtained by combining multiple datasets obtained from experiments~\cite{economides2021hierarchical,amoudruz2022}.

\subsection{Artificial Bacterial Flagella}

The \ac{ABF} is represented as a rigid body with a magnetic moment $\mathbf{m}$.
The \ac{ABF} is discretized with a set of frozen particles and a surface, similar to the representation of walls.
The magnetic moment, frozen particles and surface of the \ac{ABF} are constant in the frame of reference of the body, and the velocity of the particles follow that of the rigid body.
\Ac{DPD} particles are bounced-back from the surface of the \ac{ABF} and their change in velocity is converted into the \ac{ABF} momentum.
Furthermore, the \ac{DPD} particles interact with the frozen particles via \ac{DPD} interactions.
The sum of these interactions are accumulated into a force and torque,
\begin{equation*}
\mathbf{F}_\text{DPD} = \sum\limits_{i=1}^N \mathbf{f}_i, \;\;\;\;
\mathbf{T}_\text{DPD} = \sum\limits_{i=1}^N (\mathbf{r}_i - \mathbf{R}) \times \mathbf{f}_i,
\end{equation*}
where $\mathbf{r}_i$ are the positions of the frozen particles, $\mathbf{R}$ is the center of mass of the \ac{ABF} and, $\mathbf{f}_i$ the forces acting on these particles.
The position of the center of mass, the orientation, and the linear and angular momentum of the \ac{ABF} are then evolved according to Newton's law of motion.
In addition, the \ac{ABF} is subject to the magnetic torque $\mathbf{T}_\text{magn} = \mathbf{m} \times \mathbf{B}$, where $\mathbf{B}$ is the external magnetic field.

\section{Implementation and Parallelization strategies}

The methods described in the previous section require three types of objects: single particles that are the discretization unit of fluids; membranes, composed of a fixed amount of particles connected through a triangle mesh; rigid objects, modeling \acp{ABF}, composed of particles that are connected to the frame of reference of the rigid object.
When parallelizing such systems, both \Lmp and \Mir take a domain decomposition approach, where each MPI rank is assigned to a rectangular cuboid region, called subdomain, of the simulation space.
Objects that belong to a subdomain are stored in the memory of the associated rank.
In the case of fluid particles, the notion of ``belonging'' to a subdomain is well defined and is handled similarly in both \Lmp and \Mir.
In contrast, membrane particles might be assigned to different ranks, depending on the data structure.

In \Mir, particles of the same membrane are all stored on the rank associated with the subdomain that contains the center of mass of the membrane.
This choice results in a straightforward implementation of the membrane force computation, as all particles are local to this rank and thus do not require to be communicated.
However, this approach leads to additional communication related to interactions between membranes and other objects.
We note that the communication overlaps with computation as shown in Ref.~\cite{alexeev2020mirheo}, resulting in an excellent weak scalability.
Similarly, in \Mir, all particles in a single rigid object are stored on the same rank.
Compared to splitting particles between ranks, this approach simplifies the implementation of the time-stepping procedure of rigid objects.

In \Lmp, particles are distributed across subdomains according to their position, regardless of the objects they are associated with.
This results in membrane or \acp{ABF} particles that can be split into multiple ranks.
Thus, additional data structures were introduced to maintain the connectivity of these particles.
The standard approach provided by \Lmp, consisting of explicitly storing pairs, angles, and dihedral, would be sufficient to compute membrane forces.
However, these data structures are wasteful in terms of memory as the membranes all have the same connectivity.
In addition, these structures make the computation of the area and volume of membranes, required for the computation of the membrane forces, cumbersome and poorly scalable if not carefully implemented.
In this section, we outline the key implementation details that were essential to retaining scalability in \Lmp.
We remark that these implementation details aim to reduce the communication cost between nodes, and are thus beneficial for both traditional \ac{HPC} and cloud platforms.

\subsection{Implementation of Red Blood Cell membranes}

Due to the implementation details of the domain decomposition algorithm, particles of the same membrane may be distributed among several processors.
The computation of the membrane forces thus requires communication between ranks.
We distinguish two types of communications.
First, particles are communicated to neighboring ranks to compute the forces on bonds, triangles, and dihedrals that are local to the membrane, e.g. shear and bending forces.
We reuse the implementation in \Lmp that exchanges particles that are within a cutoff radius from each rank.
An important remark is that the connectivity of all membranes is the same, and thus we can easily reconstruct triangles and bonds by communicating the index of the membrane particles, with no extra communication required for the connectivity.
Using the \Lmp approach of storing all angles and bond indices explicitly would result in a much larger memory footprint and would be redundant in this case.
Second, the penalization terms (\cref{eq:penalization:A,eq:penalization:V}) require the total area and volume of each cell.
These quantities are computed in two steps.
First, each rank computes the partial areas and signed volumes associated with each triangle, from the local particles and those coming from neighboring ranks within a prescribed cutoff.
Second, we communicate the partial sums of the areas and volumes with neighboring ranks and sum up the results in the process, resulting in the total area and volume of the cells on each rank.
These communication patterns do not involve any all-to-all collective calls, resulting in better scaling over many ranks.
This communication pattern assumes that membranes are not larger than one subdomain length, i.e. membranes are stored in at most two consecutive ranks along each direction.
In practice, subdomains are larger than the size of one membrane, and thus this assumption is not restrictive.
An alternative approach relies on all-to-all communications, summing up partial volumes and areas of all membranes of the system.
We will demonstrate in \cref{se:lmp:scaling} that this alternative approach shows poor scaling compared to the method we have described above.

\subsection{Implementation of Artificial Bacterial Flagella}
\label{se:parallelization:ABF}

The implementation of \acp{ABF} follows the approach of rigid fix in \Lmp.
This approach attaches a rigid object index to each particle in the domain and a shared state for each object (center of mass, orientation, linear, and angular velocities) stored on every rank of the simulation.
This approach allows for simulating rigid objects that are larger than the subdomain associated with one rank.
However, the implementation relies on all-to-all communication patterns that do not scale well with the number of objects and the number of ranks in the simulations when the rigid object is much smaller than the simulation domain.
Since the \ac{ABF} has a size of the order of one single \ac{RBC}, only a few ranks need to update this \ac{ABF}.
We modified the implementation of the rigid fix in \Lmp so that only those ranks are involved in the communication, improving the scaling efficiency of the program (see \cref{se:lmp:scaling}).
The set of ranks involved in the update of the \ac{ABF} changes as the \ac{ABF} moves within the domain.
We update this set of ranks at a given frequency that is much lower than the time-stepping frequency but high enough to keep the number of ranks in this set relatively low.
This approach substantially reduces the number of all-to-all communications, and thus the scalability of the program is better than the original approach used in \Lmp, as we will see in \cref{se:lmp:scaling}.

\section{Description of the compute environments}

We use two platforms in this study: \ac{GCP} and Piz Daint at the CSCS supercomputing center.

The cloud benchmarks are run on an SLURM cluster running on \ac{GCP}.
The \ac{GPU} benchmarks are performed on a partition built out of n1-highmem-16 nodes, each equipped with 4 Nvidia T4 \acp{GPU}.
The \Lmp benchmarks use c2-standard-60 nodes, each equipped with a 60-core Intel\textsuperscript{\textregistered} Xeon\textsuperscript{\textregistered} CPU, running at a base frequency of $\SI{3.10}{\giga\hertz}$.
The cluster was built using Google's ``HPC Toolkit''~\cite{GoogleHPCToolkit}, with the addition of a \ac{GCP} Filestore instance for persistent data storage.
The software was installed using SPACK package managers~\cite{gamblin2015spack} and includes MPICH, CUDA, cmake and HDF5 libraries.

The Piz Daint supercomputer has two partitions.
The \ac{GPU} partition is a Cray X50 system with Intel\textsuperscript{\textregistered} Xeon\textsuperscript{\textregistered} E5-2690 v3 CPUs, running at a base frequency of $\SI{2.60}{\giga\hertz}$ (12 cores, 64GB RAM) and NVIDIA\textsuperscript{\textregistered} Tesla\textsuperscript{\textregistered} P100 16GB.
The multicore partition is a Cray X40 system with two Intel\textsuperscript{\textregistered} Xeon\textsuperscript{\textregistered} E5-2695 v4 running at a base frequency of $\SI{2.10}{\giga\hertz}$ (2 x 18 cores, 64/128 GB RAM).
The interconnect configuration is based on Aries routing and communications ASIC, with Dragonfly network topology.

\section{Performance of \Mir on the cloud}
\label{se:mir:scaling}

\begin{figure}
  \centering
  \includegraphics{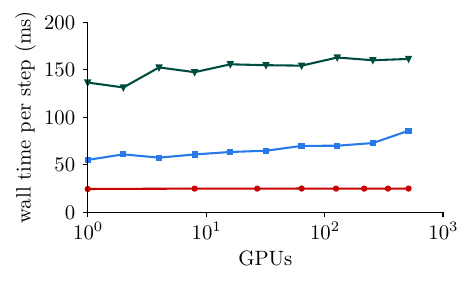}
  \includegraphics{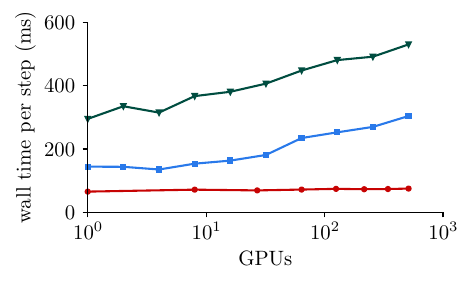}
  \caption{
    Weak scaling of \Mir on the cloud with 4 T4 GPUs per node and on Piz Daint with one P100 GPU per node.
    \key{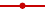}: Piz daint, $(96r_c)^3$ volume per GPU. \key{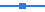}: GCP, $(96r_c)^3$ volume per GPU. \key{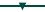}: GCP, $(128r_c)^3$ volume per GPU.
    Top: DPD particles only.
    Bottom: Bulk blood.
  }
  \label{fig:mirheo:cloud:weak}
\end{figure}

We perform scaling experiments on the cloud with tightly coupled simulations using \Mir.
\Cref{fig:mirheo:cloud:weak} shows the weak scaling of pure bulk \ac{DPD} solvent with a number density $n_d = 10r_c^{-3}$, where $r_c=1$ is the cutoff radius of the \ac{DPD} interactions.
The weak scaling experiment was repeated for subdomain volumes of $(96r_c)^3$ and $(128r_c)^3$ per \ac{GPU} (about 8.8 and 20.1 million particles per \ac{GPU}, respectively).
\Mir has a weak scaling efficiency above 80\% for a subdomain volume $(128r_c)^3$ per \ac{GPU} up to 512 \acp{GPU} (128 nodes).
In comparison, simulations on Piz Daint exhibit a weak scaling efficiency above 98\% up to 729 \acp{GPU}, due to the better bandwidth between nodes.

Similarly, we study the weak scaling of a suspension of \acp{RBC} at hematocrit $\Ht=30\%$ on the same platforms (\cref{fig:mirheo:cloud:weak}).
The weak scaling efficiency is lower than that obtained from pure \ac{DPD} bulk simulations, but remains above 50\% despite the complexity of the simulations.
The degradation of the weak scaling efficiency compared to the previous case originates from extra communication due to the bounce back of the particles against the membrane of the \acp{RBC}.
This operation cannot be overlapped with any computation.
In addition, the size of the messages is larger due to the exchange of membrane particles between neighboring ranks.

\begin{figure}
  \centering
  \includegraphics{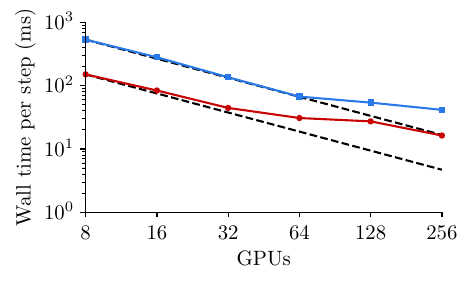}
  \includegraphics{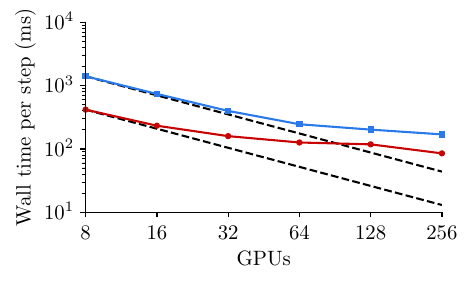}
  \caption{
    Strong scaling of \Mir on the cloud with 4 T4 GPUs per node.
    \key{111}: $(256r_c)^3$ volume, \key{122}: $(384r_c)^3$ volume, \key{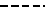}: ideal.
    Top: DPD particles only.
    Bottom: Bulk blood.
  }
  \label{fig:mirheo:cloud:strong}
\end{figure}

\Cref{fig:mirheo:cloud:strong} shows the strong scaling performance of \Mir on the cloud for a suspension of \ac{DPD} particles and the more complex scenario of blood suspension at $\Ht = 30\%$.
In the pure \ac{DPD} case, the speedup is almost ideal from 8 to 64 \acp{GPU}, and degrades for more than 128 \acp{GPU}.
In the case with blood cells, the speedup is close to ideal in the same range as the \ac{DPD} case only for the larger domain size of $(384 r_c)^3$, while the smaller domain size degrades above 32 \acp{GPU}.
As expected, the speed-up is closer to ideal for larger domain sizes, since there is more work per \ac{GPU}.
The case including blood cells shows worse strong scaling than the pure \ac{DPD} case, caused by the higher amount of communication required to exchange membrane information, the larger number of messages due to the different types of solvent particles (cytosol and plasma), and the additional communication due to bounce-back on membranes.

\section{Performance of \Lmp on the cloud}
\label{se:lmp:scaling}

\begin{figure}
  \centering
  \includegraphics{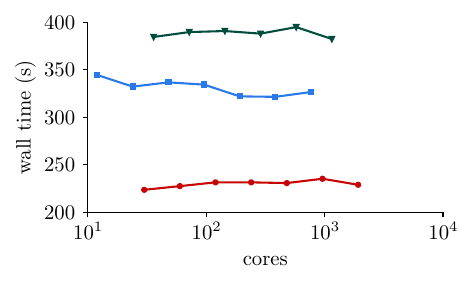}
  \includegraphics{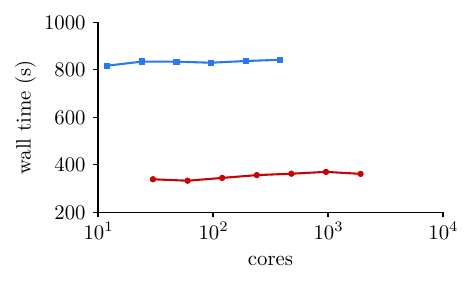}
  \caption{
    Weak scaling performance of \Lmp.
    Top: DPD only, $(128r_c)^3$ volume per node, \key{111}: GCP c2-standard-60 partition, 30 cores per node, \key{122}: Piz Daint GPU-partition 12 cores per node, \key{133}: Piz Daint multi-core partition, 36 cores per node.
    Bottom: Blood suspension with $30\%$ hematocrit, $(128r_c)^3$ volume per node, \key{111}: GCP c2-standard-60 partition, 30 cores per node, \key{122}: Piz Daint GPU-partition 12 cores per node.
  }
  \label{fig:lammps:cloud:weak}
\end{figure}

\begin{figure}
  \centering
  \includegraphics{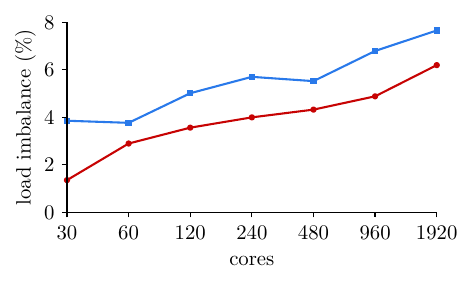}
  \caption{
    Weak scaling of \Lmp on the C2 partition: load imbalance for a volume of $(128r_c)^3$ per node.
    \key{111}: Pure DPD particles, \key{122}: Blood suspension with $30\%$ hematocrit.
  }
  \label{fig:lammps:cloud:weak:imbalance}
\end{figure}

We examine the scaling capabilities of \Lmp on the C2-standard partition.
\Cref{fig:lammps:cloud:weak} shows the weak scaling efficiency of \Lmp for pure \ac{DPD} dynamics and for a $30\%$ hematocrit blood suspension.
The simulations of pure \ac{DPD} solvent are set in a periodic domain, with a number density of $n_d = 10 r_c^{-3}$, where $r_c$ is the cut-off radius of the \ac{DPD} interactions.
Each node consists of 30 physical cores per subdomain of volume $(128r_c)^3$.
The suspension of \acp{RBC} has a hematocrit of $30\%$.
The \ac{RBC} membranes are composed of 642 vertices and their size corresponds to an equivalent radius of $R_A = 3 r_c$, with $R_A = \sqrt{A_0 /(4\pi)}$. $A_0$ is the membrane area.
Each simulation is repeated five times, and the minimum wall time over 100 steps is recorded.
The scaling efficiency remains above 95\% for pure solvent and above 90\% for the more complex case of a blood suspension, up to 1920 cores (64 nodes).
\Lmp has a similar weak scaling efficiency on two partitions of the Piz Daint supercomputer, although the CPU frequencies are lower and thus the time to solution is higher on this platform.
\Cref{fig:lammps:cloud:weak:imbalance} shows the load imbalance for both pure \ac{DPD} dynamics and blood suspension on \ac{GCP}.
The load imbalance was measured by enabling the \textit{timer full sync} option of \Lmp, which synchronizes all ranks after each phase of the simulation time steps, by performing a collective barrier call.
The synchronization time was measured and normalized by the simulation time, forming the load imbalance.
We note that this quantity is an upper bound of the load imbalance, as this also includes the time to perform the collective barrier calls.
Each simulation was run five times and the minimal load imbalance was reported.
The load imbalance remains below 8\% for all simulations performed and is, as expected, lower for the pure \ac{DPD} solvent than for the blood suspension.
Furthermore, the load imbalance deteriorates as the number of cores increases.

\begin{figure}
  \centering
  \includegraphics{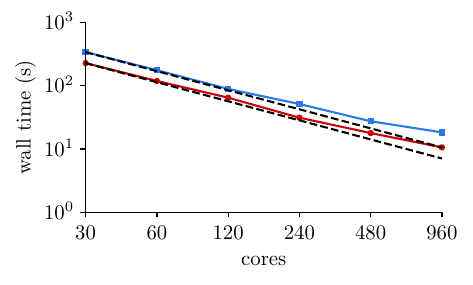}
  \caption{
    Strong scaling of \Lmp on the C2 partition with $(128r_c)^3$ volume.
    \key{111}: DPD particles only.
    \key{122}: Bulk blood at $30\%$ hematocrit.
    \key{200}: ideal.
  }
  \label{fig:lammps:cloud:strong}
\end{figure}

Finally, \cref{fig:lammps:cloud:strong} shows the strong scaling performance of \Lmp on the C2 partition.
The simulation consists of the same cases used during the weak-scaling study, in a domain of fixed volume $(128r_c)^3$.
The simulations show excellent strong scaling as the speed-up remains close to ideal up to 480 cores (16 nodes), for both cases considered in this section.
Beyond 480 cores, the strong scaling efficiency decreases.
This is likely due to the load imbalance, as seen in \cref{fig:lammps:cloud:weak:imbalance}, and increased communication latency, which becomes more significant compared to the computation load as the number of cores increases.
To further improve strong scaling efficiency, strategies include hybrid MPI/OpenMP approach to reduce the number of messages and asynchronous communication to better overlap communication with computation.

\begin{figure}
  \centering
  \includegraphics{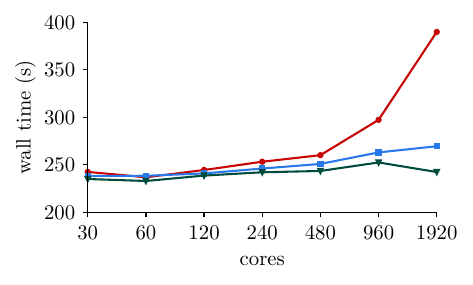}
  \caption{
    Weak scalability of a $(128 r_c)^3$ blood simulation (similar to \cref{fig:lammps:cloud:weak}) with different data structures to store the RBC membrane connectivity.
    \key{111}: Default \Lmp approach, \key{122}: local reconstruction of connectivity with collective communication, \key{133}: local reconstruction of connectivity with neighboring communication.
  }
  \label{fig:lammps:cloud:blood:improvement}
\end{figure}

\Cref{fig:lammps:cloud:blood:improvement} demonstrates the benefits of not explicitly storing the \acp{RBC} connectivity for each membrane.
Using \Lmp structures (bonds, angles and dihedrals) leads to larger messages and a larger memory footprint, which is unnecessary as all \ac{RBC} membranes have the same connectivity.
Instead, our approach reconstructs the connectivity of the triangle mesh on each rank at every iteration from the particle global indices without communicating them.
This approach decreases the message sizes between ranks and the memory footprint of the simulations.
We tested the different implementations on a blood suspension with $\Ht=30\%$ and a volume per node (30 cores) of $(128 r_c)^3$, with membranes of effective radius $R_A=3r_c$ and 642 vertices per membrane.
It is clear from \cref{fig:lammps:cloud:blood:improvement} that the new approach scales much better than using the \Lmp structures.

The calculation of membrane forces requires one to calculate the volume and area of each membrane at each time step, as shown by \cref{eq:penalization:A,eq:penalization:V}.
Due to the domain partitioning, \acp{RBC} may be split between multiple ranks, therefore the initial implementation used collectives to calculate the area and volume of cells.
Since \acp{RBC} have a finite size, usually only a few ranks are involved in the computation of each \ac{RBC} area and volume.
Thus, we exchange the partial volume and areas of the membranes only with the ranks containing parts of the membranes.
This approach further improves the weak scaling of the blood suspension simulations compared to using an all-to-all communication pattern, as shown on \cref{fig:lammps:cloud:blood:improvement}.

\begin{figure}
  \centering
  \includegraphics{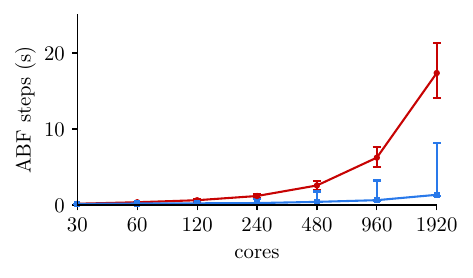}
  \caption{
    Time spent evolving the \ac{ABF} against number of cores while increasing the domain size proportionally,  using \textit{full sync} timer option in \Lmp.
    The curve denotes the average time per core, and the error bars show the minimal and maximal time among all cores.
    \key{111}: Default \Lmp implementation, \key{122}: Sub-communicator implementation.
  }
  \label{fig:lammps:cloud:abf:weak}
\end{figure}

\Cref{fig:lammps:cloud:abf:weak} shows the time spent updating a single \ac{ABF} in a pure \ac{DPD} solvent over 1000 steps.
The volume of the domain increases proportionally to the number of cores, and we use a volume of $(32 r_c)^3$ per node (30 cores).
The default approach used in \Lmp uses collective calls, resulting in poor scalability.
We improved the communication pattern by restricting collective communication calls to only the ranks in the vicinity of the \ac{ABF} particles (see \cref{se:parallelization:ABF}).
This approach results in much better scalability than the original \Lmp approach.
The measurements were performed with the synchronized and nonsynchronized timer options in \Lmp, and no major differences were found between the two, suggesting that the load imbalance between ranks in earlier stages does not influence these results.
We note that the maximum time spent per rank for this stage of the simulation is not negligible, as this timing takes into account the computation part of the update.
However, the minimum and average time across all ranks remain very low with the improved communication pattern, even when the number of ranks is large.

We remark that while \Lmp has better weak scalability than \Mir in the above tests, the time to solution of \Mir is much lower.
On the pure \ac{DPD} dynamics in a volume of $(128r_c)^3$ per T4 \ac{GPU}, \Mir exhibits a wall time of $\SI{130}{\milli\second}$ per time step on one \ac{GPU} while \Lmp takes about $\SI{2.23}{\second}$ per step on one node (17 times longer).
Thus, the communication time represents a larger part of the simulations in \Mir than in \Lmp, possibly explaining the lower weak scaling efficiency in this specific case.

\begin{table}[h!]
\centering
\begin{tabular}{lllll}
 & \multicolumn{2}{c}{\textbf{Fluid}} & \multicolumn{2}{c}{\textbf{Blood}} \\
N & Palabos & LAMMPS & Palabos & LAMMPS \\
\hline
1   & --   & --   & 1.00 & 1.00 \\
2   & 1.00 & 1.00 & 0.89 & 0.97 \\
4   & 0.96 & 0.92 & 0.74 & 0.95 \\
8   & 0.90 & 0.95 & 0.57 & 0.83 \\
16  & 0.81 & 0.83 & --   & 0.77 \\
32  & 0.74 & 0.70 & --   & 0.58 \\
64  & 0.69 & --   & --   & --   \\
128 & 0.57 & --   & --   & --   \\
\end{tabular}
\caption{Comparison of strong scaling efficiency between Palabos (from AWS benchmarks~\cite{Tang2023AWS})
and LAMMPS (this work) for fluid and blood simulations, against the number of instances $N$.}
\label{tab:palabos_lammps}
\end{table}

Finally, we compare the strong scaling efficiency of \Lmp with that of Palabos~\cite{kotsalos2021palabos}.
Palabos was benchmarked on AWS EC2 Hpc7g instances with 64 cores per instance~\cite{Tang2023AWS} for two representative cases: a pure fluid system, with 1 billion grid cells, and a blood flow case with 476 \acp{RBC} and 95 platelets.
As shown in \cref{tab:palabos_lammps} Palabos and \Lmp exhibit comparable strong scaling efficiency for the fluid case, despite Palabos being run at a larger problem size (1 billion grid cells versus 20 million particles in our simulations).
In the more complex case of blood, where we implemented algorithmic improvements for the membrane communication patterns, \Lmp achieved higher strong scaling efficiency.
This improvement is also supported by the larger problem size considered in this work (7\,489 \acp{RBC} compared to 476 in the Palabos study), which provides more computational work per core and thus better scaling efficiency.

\section{Cost efficiency}

In this section, we compare the cost efficiency of the simulations benchmarked in \cref{se:mir:scaling,se:lmp:scaling}.
Specifically, we evaluate the average number of simulation steps achieved per unit cost, as summarized in \cref{tab:cost:efficiency}.
In all cases, Piz Daint achieves a higher number of simulation steps per unit cost than \ac{GCP}.
However, the gap is relatively modest: \ac{GCP}'s cost efficiency is within a factor of 2 compared with Piz Daint.

It is important to note that the reported cost on Piz Daint (CHF 0.40 per node-hour) is heavily subsidized, as is typical for publicly funded supercomputing centers.
This cost does not reflect the full operational expenses of the system, which include infrastructure, personnel, and energy costs.
In contrast, the prices on \ac{GCP} represent full commercial rates, including overheads and profit margins.
From this perspective, the cost efficiency of cloud resources appears competitive, especially considering the flexibility, scalability, and accessibility they provide.

\begin{table}
  \centering
  \begin{tabular}{lrr}
    \textbf{Case} & \textbf{Piz Daint X50} & \textbf{GCP} \\\hline
    \Mir, DPD, $(96 r_c)^3$ & 306\,115 & 196\,818 \\
    \Mir, blood, $(96 r_c)^3$ & 115\,024  & 72\,439 \\
    \Lmp, DPD, $(128 r_c)^3$ & 2\,177 & 1\,028 \\
    \Lmp, blood, $(128 r_c)^3$ & 918 & 678
  \end{tabular}
  \caption{Cost efficiency (in number of simulation steps per dollar) of the DPD and blood cases obtained with \Lmp and \Mir on both the Piz Daint supercomputer and GCP.}
  \label{tab:cost:efficiency}
\end{table}

\section{ABF in a retinal capillary network}

The benchmarks presented in the previous sections demonstrate the weak and strong scaling capabilities of \Lmp on \ac{GCP} for homogeneous, isotropic cases (periodic \ac{DPD} fluids and blood suspensions).
However, more realistic systems include complex geometries and large-scale simulations, possibly coupled with additional physics, and complex flows.
In this section, we demonstrate that our approach generalizes to such cases by simulating the evolution of an \ac{ABF} swimming in a retinal capillary network.

\begin{figure}
  \centering
  \includegraphics{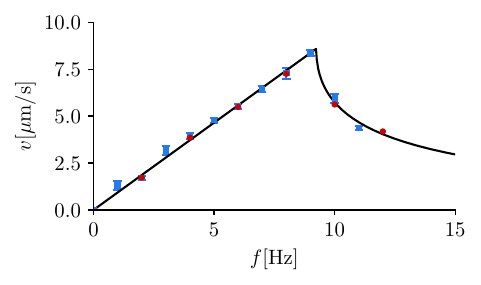}
  \caption{Swimming speed of an ABF in a viscous fluid against the rotation frequency of the external magnetic field.
    \key{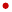}: DPD simulations;
    \key{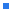}: experimental measurements from Mhanna et al.~\cite{mhanna2014artificial};
    \key{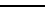}: fit to the experimental data with the ODE model explained in refs.~\cite{schamel2013chiral,vach2013selecting}.}
  \label{fig:abf:validation}
\end{figure}

First, to validate the \ac{ABF} model, we simulate a single swimmer in a viscous fluid with zero background velocity.
The domain is periodic with large enough dimensions to avoid interactions between the \ac{ABF} and its periodic images.
A uniform, rotating magnetic field with constant frequency $f$ propels the swimmer to an average velocity $v$.
The swimming speed of \acp{ABF} in such conditions has been measured experimentally~\cite{mhanna2014artificial} and explained with an ODE model~\cite{schamel2013chiral,vach2013selecting}.
\Cref{fig:abf:validation} demonstrates that the \ac{DPD} model used in this work has an excellent agreement with both theory and experiments: below a critical frequency of the magnetic field, the swimming speed of the \ac{ABF} increases linearly; above the critical frequency, however, the time-averaged speed of the \ac{ABF} decreases due to the bounded magnetic torque~\cite{schamel2013chiral,vach2013selecting,amoudruz2022phd}.

\begin{figure*}
  \centering
  \includegraphics[width=0.8\textwidth]{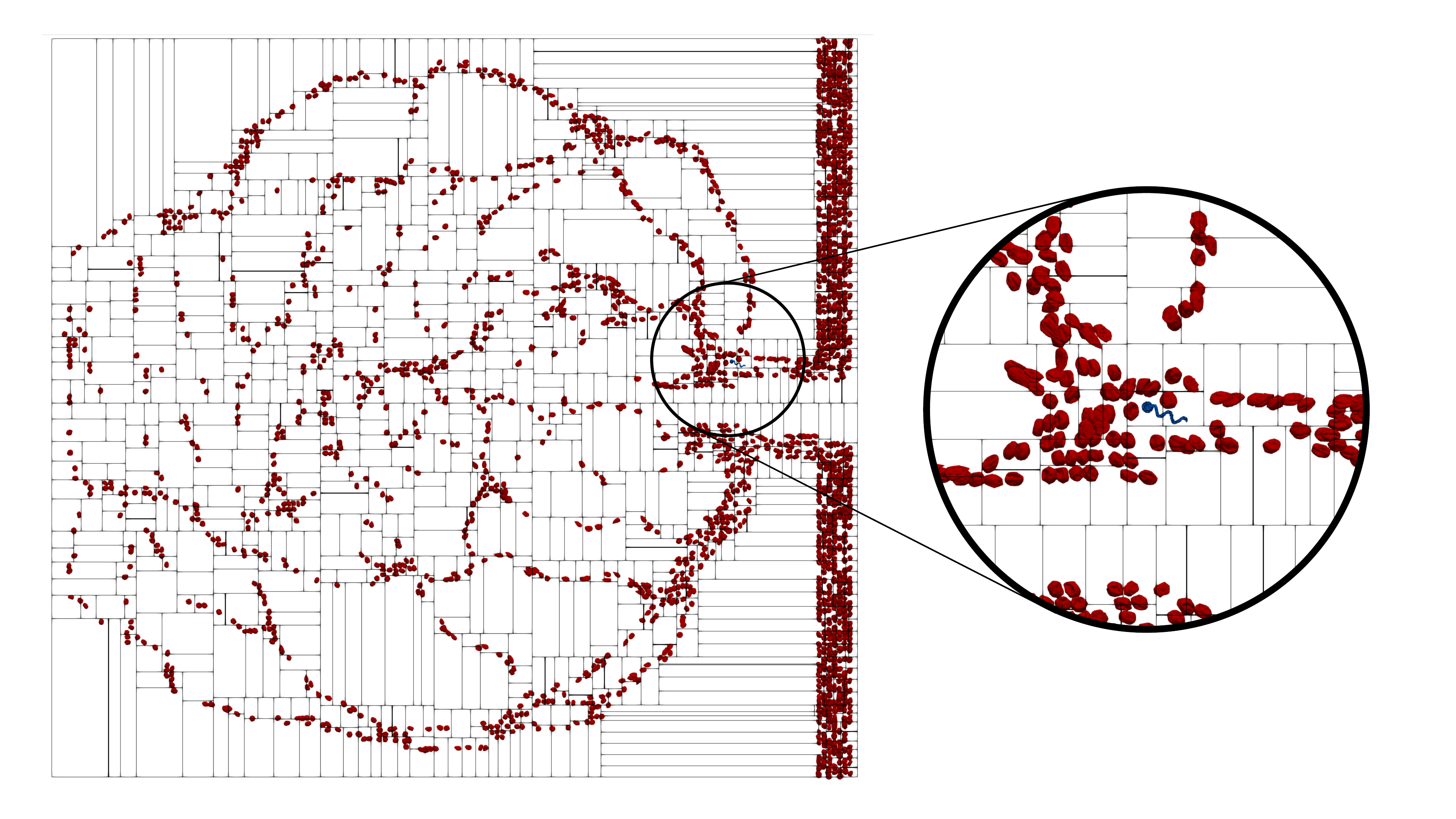}
  \caption{
    Simulation of an ABF swimming through retinal capillaries.
    Capillary walls and solvent particles are not shown, for clarity.
    The rectangular boundaries represent the domain decomposition (one MPI rank per subdomain).
  }
  \label{fig:capillaries:snapshot}
\end{figure*}

We now extend these results to the case of an \ac{ABF} in a retinal capillary network.
The geometric representation of the capillaries is reproduced from a fundus image of a retinal capillary network~\cite{ghassemi2015}.
The simulations are carried out with \Lmp with a \textit{tiled} domain decomposition to ensure load balancing between ranks.
The \ac{ABF} is guided by an external magnetic field, controlled by a \ac{RL} agent.
The \ac{RL} agent was trained as described in~\cite{amoudruz2025optimalnavigationmagneticartificial} from a reduced-order model of the simulations.
The policy of the \ac{RL} agent maps the center of mass of the swimmer to the desired swimming direction $\mathbf{p}$.
We thus apply a magnetic field rotating in a plane perpendicular to the swimming direction,
\begin{equation*}
  \mathbf{B}(t) = R_x(\mathbf{p}) B
  \begin{pmatrix}
    \cos \omega t \\
    \sin \omega t \\
    0
  \end{pmatrix},
\end{equation*}
where $R_x(\mathbf{p})$ is the rotation that transforms the vector $\mathbf{e}_x$ into the swimming direction $\mathbf{p}$ with axis of rotation $\mathbf{e}_x \times \mathbf{p}$, and $\mathbf{e}_x$ is the principal axis of the non-rotated \ac{ABF}.
The magnetic field is adapted at every time step of the simulation.

The largest capillary has a radius of $\SI{15}{\micro\meter}$.
A force field is applied to every particle to mimic a pressure gradient.
The magnitude of this force is set to achieve a mean inflow velocity of $\SI{1}{\milli\meter\per\second}$.
The hematocrit is set to 20\%.
These numbers are typical for blood flow in capillaries~\cite{fung1998}.
The magnetic moment has a magnitude of $m=\SI{1e-11}{\newton\meter\per\tesla}$ and the external magnetic field has a magnitude of $B=\SI{3}{\milli\tesla}$ and a rotation frequency $f=\SI{1}{\kilo\hertz}$.
The resolution of the \ac{DPD} solvent corresponds to a number density of $10 r_c^{-3}$, with $R_A = 4r_c$, resulting in approximately 90 million particles.
We employ the C2 partition with 30 physical cores per node, on a total of 2490 cores (83 nodes).
\Cref{fig:capillaries:snapshot} shows a snapshot of the \ac{ABF} swimming through the complex network of capillaries.
We note that in this case the tiled domain decomposition is crucial to saving computational resources, as the geometry is relatively sparse.

\section{Limitations}

In this work, we have demonstrated that cloud platforms are well suited for large-scale, tightly coupled simulations.
The two software frameworks tested show excellent weak and strong scalability, comparable to traditional supercomputers.
However, these benchmarks were performed on a single rack, where the communication cost is minimal.
Scaling up to simulations spanning multiple racks, network islands, or even datacenters  would inevitably increase the communication latency by one or two orders of magnitude, while reducing bandwidth.
In turn, the communication-to-computation ratio increases, leading to a drop in scaling efficiency.
This is a general limitation of cloud infrastructures compared to specialized HPC clusters with low-latency interconnects such as InfiniBand.
To mitigate these effects, hybrid parallelization strategies such as MPI/OpenMP can reduce the number of inter-node messages and help maintain efficiency by minimizing latency overhead.
Additionally, careful resource placement and use of cloud-specific optimizations (e.g., placement policies) can further alleviate communication bottlenecks.

The simulations presented in this work are based on state-of-the-art models of blood at the microscale.
Nonetheless, some idealizations were necessary.
For instance, we have neglected the presence of white blood cells and platelets, which may affect the hydrodynamics in capillary networks.
In the retinal microvasculature simulations, vessel walls were modeled as no-slip and no-through boundaries.
However, in vivo, the endothelial glycocalyx layer and cellular permeability enable some degree of fluid and molecular exchange with surrounding tissues.
These simplifications may influence the interaction of \acp{RBC} and \acp{ABF} with the vessel boundaries.
While recent studies have started to incorporate such effects~\cite{jiang2021understanding}, their influence in the context of \acp{ABF} remains largely unexplored and requires further investigations.

\section{Summary}

We have demonstrated that the cloud is well suited for large-scale, complex, tightly coupled scientific simulations.
We have tested the scalability of two programs on different cloud architectures.
\Mir shows excellent weak scalability up to 512 \acp{GPU}.
Furthermore, excellent scalability was achieved with the widely used software \Lmp for bulk \ac{DPD} simulations.
\Lmp was further extended to simulate blood suspensions at physiological hematocrits and the swimming of \acp{ABF} within complex retinal capillaries.
By reducing the size of point-to-point communication and minimizing collective communications, we were able to retain weak scaling efficiency above 90\% for up to 1920 cores.
This was achieved by improving data-structures of the membrane connectivity, and by adapting communication patterns between nodes accordingly.
We have shown that \Lmp can simulate an \ac{ABF} swimming in a complex network of capillaries that was reconstructed from fundus images of retinal capillaries, which paves the way for democratizing access to critical simulations for personalized medicine.
The configuration of the cloud platform provides an environment similar to that of traditional supercomputers.
This methodology is thus reusable for any other large-scale scientific applications.
These results highlight cloud computing as a viable platform that complements traditional HPC by providing accessible and competitive performance for complex, large-scale, tightly coupled simulations.

\section*{Acknowledgment}

The authors thank Dan Speck, Matthew Cruger, and Mark Heil for helpful discussions and cloud support.

\bibliographystyle{elsarticle-num}
\bibliography{refs}

\end{document}